\documentclass[12pt,preprint]{aastex}

\slugcomment{Submitted to The Astrophysical Journal}

\shorttitle{Speed of Gravity and Time Delay}
\shortauthors{Clifford M. Will}

\begin{document}


\title{Propagation Speed of Gravity and the Relativistic Time Delay }


\author{Clifford M. Will}
\affil{
Department of Physics, McDonnell Center for the Space Sciences, \\
Washington University, St. Louis, MO 63130}


\begin{abstract}
We calculate the delay in the propagation of a light signal past a
massive body that moves with speed $v$, 
under the assumption that the speed of propagation of
the gravitational interaction $c_g$ differs from that of light.  
Using the post-Newtonian approximation,
we consider an expansion in powers of
$v/c$ beyond the leading ``Shapiro'' time delay effect, while working to
first order only in $Gm/c^2$, and show
that the altered propagation speed of the gravitational signal has no effect 
whatsoever on
the time delay to first order in $v/c$ beyond the leading term, 
although it will have an effect to
second and higher order.  We show that the only other possible effects  
of an altered speed $c_g$ at this order arise from
a modification of the parametrized
post-Newtonian (PPN) coefficient $\alpha_1$ of the metric from the value
zero
predicted by general relativity.  Current solar-system
measurements already provide tight bounds on such a modification. 
We conclude that recent measurements of the propagation of radio
signals past Jupiter are sensitive to $\alpha_1$, but are {\em not} 
directly sensitive to the
speed of propagation of gravity. 
\end{abstract}




\section{Introduction}

The time delay of light, also known as the Shapiro effect, is one of
the observational cornerstones of general relativity.  For a light ray
passing a body of mass $m_a$ at location ${\bf x}_a$, and received by
an observer at ${\bf x}_r$, the excess delay is given by (modulo a
constant)
\begin{equation}
\Delta = -{2Gm_a \over c^3} \ln [ r_{ra} - {\bf k} \cdot {\bf x}_{ra}
] \,,
\label{eq1}
\end{equation}
where ${\bf k}$ is a unit vector in the direction of the incoming
light ray, ${\bf x}_{ra} = {\bf x}_r - {\bf x}_a$, and $r_{ra} = |
{\bf x}_{ra}|$.  If the ray passes by the body a minimum distance $d$ that is
small compared to $r_{ra}$, then 
\begin{equation}
\Delta \approx {2Gm_a \over c^3} \ln \left ( {2r_{ra} \over d^2} \right
) \,.
\end{equation}
The one-way delay amounts to  70 microseconds for a ray that grazes
the Sun and 10 nanoseconds for a ray that grazes Jupiter.  In addition to tests
of the time delay by the sun \citep{reasenberg}, the time delay has also
been studied in a number of binary pulsar systems (see for example, 
\citet{stairs}).

Advances in precision timekeeping and in VLBI have motivated the
study of the first relativistic corrections, of order $v/c$, to the
basic Shapiro formula.  These corrections 
 would be relevant for signals passing near moving bodies, such
as bodies in binary pulsar systems, or planets such as Jupiter. 
In a series of papers, Kopeikin and colleagues have analyzed
these correction effects in detail (see \citet{KopSch99} and
references therein).  They find that, to first order in
$v_a/c$, where $v_a$ is the velocity of body $a$, 
the corrected expression takes the form
\begin{eqnarray}
\Delta &=& -{2Gm_a \over c^3} \left ( 1- 
{{{\bf k} \cdot {\bf v_a}} \over c} \right ) 
\ln [ r_{ra} - {\bf K} \cdot {\bf x}_{ra} ]  
\nonumber \\
&& + O\left ( {Gm_a \over c^3} {v_a^2 \over c^2} \right )\,,
\label{eq2}
\end{eqnarray}
where
\begin{equation}
{\bf K} = {\bf k} - {1 \over c} {\bf k} \times ({\bf v_a} \times
	 {\bf k}) \,,
\label{eq3}
\end{equation}
where all quantities are to be evaluated at the same moment of time,
say, the time of reception of the ray.

For a  light ray passing close to Jupiter, these corrections are small,
of order picoseconds or less, 
but may be detectable with the latest VLBI techniques
\citep{KopFom02}.

Recently, however Kopeikin has argued that, because of the motion of a
body such as Jupiter during the passage of the light ray, the fact
that the gravitational interaction is not instantaneous should have an
effect on the time delay \citep{Kop01}.  
To study this phenomenon, he allowed the
speed of gravity to be $c_g$, where $c_g \ne c$; this would modify the
retardation of the gravitational fields used to calculate the time
delay. 
He argued that Eq. (\ref{eq2}) would be modified by replacing $c$ by
$c_g$ in the prefactor  $1- {\bf k} \cdot {\bf v_a}/c$ in Eq.
(\ref{eq2}), and in the
velocity-dependent term in Eq. (\ref{eq3}) \citep{Kop02}.
\citet{KopFom02} then proposed a test of the propagation
speed of gravity via VLBI measurements in the fall of 2002.

This result goes against intuition.  First, while the retardation of
gravity surely has effects, one does not expect to find effects {\em odd} in
powers of $v/c$ until the leading effects of gravitational radiation
reaction occur, far beyond the order of the $v$-dependent effects
claimed by Kopeikin.  Second, the time delay is controlled dominantly
by the Newtonian gravitational potential $U$; if this is a retarded
potential with speed $c_g$, it would be given by
\begin{equation}
U(t,{\bf x}) =\int {{\rho(t - |{\bf x} - {\bf x}^\prime |/c_g,{\bf x}^\prime)}
\over
        {|{\bf x} - {\bf x}^\prime |}}  d^3x^\prime \,,
\label{Uretarded}
\end{equation}
where $\rho$ is the density of matter.
Expanding the retardation in the near zone ($|{\bf x} - {\bf x}^\prime
| \ll$ a characteristic wavelength of gravitational waves), where the
time delay predominantly occurs, we obtain
\begin{eqnarray}
U &=& \int {{\rho(t,{\bf x}^\prime)} \over
        {|{\bf x} - {\bf x}^\prime |}}  d^3x^\prime
- {1 \over c_g}
  {\partial \over {\partial t}} \int \rho(t,{\bf x}^\prime)
d^3x^\prime
\nonumber \\
&& + {1 \over 2c_g^2}{\partial^2 \over {\partial t^2}}
\int \rho(t,{\bf x}^\prime) |{\bf x} - {\bf x}^\prime | d^3x^\prime +
\dots
\,,
\end{eqnarray}
The second term would have been of order $v/c_g$ relative to the leading
term, and might have contributed to the time delay, but it
vanishes because of the conservation of
mass.  
The effect of $c_g$ first appears in the third term, which is of order
$(v/c_g)^2$ relative to the first.   This, rather than first order in
$v/c_g$, is what intuition dictates would be the first level at
which retardation-dependent effects occur.  This  
argument was invoked implicitly in the paper of \citet{asada}.

Another argument against such $v/c_g$ dependent effects comes from
electrodynamics.  It is a well-known fact that the electric field of a
uniformly moving charge points not toward the retarded position of the
charge, but toward the {\it present} position (see, for example, \S
11.10 and Exercise 11.17 of \citet{Jackson}).  In gravity, because
acceleration involves $Gm/r \sim v^2$, working to first order in $v/c$
is equivalent to assuming a uniformly moving body, and thus a similar
conclusion holds: to order $v/c$, retardation should have no observable
effects.\footnote{A similar conclusion was reached by \citet{carlip} in the
context of binary motion, in response to a proposed bound on the speed of
gravity by \citet{vanflandern}.}

These considerations have motivated us to check Kopeikin's claim with
a careful calculation of the time delay to the same order.  We have
assumed that, to the post-Newtonian order required for the
calculation, the metric can be described by a matter gravitational
potential $U$ and a
current gravitational potential $V^i$, both of which are
assumed to be
retarded potentials using a speed $c_g$, as in Eq. (\ref{Uretarded}).  
The terms in the metric also
are multiplied by suitable parametrized post-Newtonian (PPN) parameters 
to reflect the possibility that, whatever theory one is considering
with a different speed of gravity, it will necessarily {\em not} be
general relativity, and thus may have different PPN parameter values 
(see \citet{tegp} for a review of the PPN framework).  In
such a theory, those PPN parameters may well depend on $c_g$, but in
the absence of a specific theory, we leave them general.  

We then calculate the time delay to order $(Gm_a/c^3)(v_a/c)$, for a
signal emitted from a distant source and received on Earth, passing
through the gravitational field of the solar system.  We carefully
treat the propagation of the light ray both through the far zone,
where the propagation time is long compared to the dynamical time of
the system, and through the near zone, where the propagation time is
short compared to the dynamical time, and where positions and
velocities can be expanded about values at some common time, such as
the time of reception on Earth.  Our
result closely parallels Eq. (\ref{eq2}):
\begin{eqnarray}
\Delta &=& -{2Gm_a \over c^3} \left ( {1+\gamma \over 2} \right )
\left ( 1- (1+\zeta)
{{{\bf k} \cdot {\bf v_a}} \over c} \right )
\ln [ r_{ra} - {\bf K} \cdot {\bf x}_{ra} ]  
\nonumber \\
&& + O\left ( {Gm_a \over c^3} {v_a^2 \over c^2} \right )\,,
\label{eq4}
\end{eqnarray}
with $\bf K$ given by Eq. (\ref{eq3}).  

The first important
observation is that 
$c_g$ appears nowhere in this expression.  Although it appears in
various intermediate expressions, it {\em cancels} in the end, to the
order of approximation shown.    
In addition to the overall
factor of $(1+\gamma)/2$, which is standard, the only change comes in
the coefficient $(1+\zeta)$ in front of the velocity-dependent term in
the prefactor.  Here $\zeta$ is a function of the PPN parameters
$\alpha_1$ and $\gamma$, given by 
\begin{equation}
\zeta = { \alpha_1 \over (2+2\gamma)}
\end{equation}
In general relativity, $\gamma=1$, $\alpha_1=0$, and we recover Kopeikin's
result, Eq. (\ref{eq2}).

Secondly, the velocity dependent corrections in Eq. (\ref{eq4}) have a
simple physical interpretation.  
Since we are working to first order in 
$v/c$, only Jupiter's rectilinear motion is relevant; its acceleration
generates effects higher order in $Gm/c^2$.  
The vector $\bf K$ is simply the spatial unit vector of the photon's
direction {\it as seen in the rest frame of Jupiter}; the correction terms
in Eq. (\ref{eq3}) are nothing but the effect of aberration.  They depend
only on $c$, which is the velocity relevant for Lorentz transformations and
the propagation of light, not on $c_g$.  Indeed, from the viewpoint of
Jupiter's rest frame,
$c_g$ is completely irrelevant, since the gravitational field is static
(again, ignoring Jupiter's acceleration).  
The velocity-dependent prefactor is a combination of a gravitomagnetic effect
(the fact that the gravitational field is not a scalar quantity, but
contains both a vector and tensorial part) and a simple Doppler effect in
transforming from Jupiter's rest frame to the barycentric frame.  

We thus conclude that the $v/c$ corrections to the Shapiro time delay
are normal 1.5PN corrections that occur when there are moving bodies,
but that they have {\em nothing} to do with the speed of propagation
of gravity, insofar as it affects the retardation of gravitational
interactions.  Furthermore, as a potential test of alternative
gravitational theories, measuring these $v/c$ terms is not promising,
because a variety of solar system measurements already constrain
$\alpha_1$ and $\gamma$ to such a degree that $|\zeta | < 4 \times 10^{-3}$
under relatively weak assumptions or $|\zeta | < 5 \times 10^{-5}$ under
assumptions that invoke so-called ``preferred-frame'' tests of the
parameter $\alpha_1$ (see
\citet{livrev} for the latest bounds on the PPN parameters).  In fact, the
VLBI measurements are sensitive mainly to the velocity dependence in
the logarithmic term, not to the prefactor.
Therefore, measurements of the propagation of radio waves past
Jupiter do not directly constrain the propagation speed of the
gravitational interaction.

The remainder of this paper provides details.  In Sec. \ref{basic} we
describe the assumptions and basic equations that go into our
calculation.  Section \ref{lienard} carries out the integrations to
find the time delay.  In Sec. \ref{discussion} we give concluding
remarks.

\section{Assumptions and Basic Equations}
\label{basic}

We assume that, whatever theory of gravity is in force, it is a {\em
metric} theory, that is it has a spacetime metric $g_{\mu\nu}$ that
governs the interactions and motions of all non-gravitational fields
and all ``test'' particles.  In particular, 
we will assume that
light rays move on null geodesics of the spacetime metric.
We will assume that atomic clocks measure proper time as given by the
invariant interval of the spacetime metric.  By choosing the units
determined by such physical measurements appropriately, we may make
the speed of light as measured by any freely falling observer unity.
Henceforth we will use units in which $G=c=1$.

We then assume that the theory of gravity has a post-Newtonian limit
that can be written in the following form:
\begin{eqnarray}
g_{00} &=& -1 + 2U + O(\epsilon^2) \,,
\nonumber \\
g_{0i} &=& -2(1+\gamma+ \alpha_1/4) V_i + O(\epsilon^{5/2}) \,,
\nonumber \\
g_{ij} &=& \delta_{ij} (1+2\gamma U) + O(\epsilon^2) \,,
\label{metric}
\end{eqnarray}
where $U$ and $V_i$ are retarded gravitational potentials given by
\begin{eqnarray}
U &=& \int {{\rho(t^\prime,{\bf x}^\prime)} \over 
	{|{\bf x} - {\bf x}^\prime |}} \delta(t^\prime - t + 
|{\bf x} - {\bf x}^\prime |/c_g) d^4x^\prime \,,
\nonumber \\
V_i &=& \int {{\rho(t^\prime,{\bf x}^\prime)v^i(t^\prime,{\bf x}^\prime)} \over
	{|{\bf x} - {\bf x}^\prime |}} \delta(t^\prime - t +
	|{\bf x} - {\bf x}^\prime |/c_g) d^4x^\prime \,,
\end{eqnarray}
where $\rho$ and $v^i$ are the density and velocity of matter, and where we
have inserted $c_g$ in the retardation expression to reflect the possible
difference between the speed of propagation of gravity and that of light
(unity).  The parameter $\epsilon$ is a bookkeeping parameter inserted to
keep track of ``orders of smallness'' of various quantities, with
$v \sim \epsilon^{1/2}$, $U \sim \epsilon$, $V_i \sim \epsilon^{3/2}$, and
so on.  

To the orders of approximation shown, the metric in Eq. (\ref{metric})
is general enough to encompass a broad class of metric theories of
gravity with a propagation speed of gravity different from unity, with
the exception of ``massive graviton'' theories
\citep{visser,grishchuk}, whose gravitational potentials also include
Yukawa-like modifications.  In the special case of $\gamma=1$,
$\alpha_1=0$ and $c_g =1$, the metric corresponds precisely to that of
linearized general relativity in harmonic gauge, to the order considered.  
There are two free PPN coefficients
$\gamma$ and $\alpha_1$; the coefficient of $U$ in $g_{00}$ has been
chosen to be unity as usual by our choice of the measured value of the
gravitational constant.   We ignore all ``preferred frame'' terms that
could appear in the PPN metric (see \citet{tegp} for discussion).

Notice that the terms ignored in each metric component
are all $O(\epsilon)$ larger than the
terms kept; no correction terms a ``half-order'' higher are present, such as
$O(\epsilon^{3/2})$ terms in $g_{00}$ or $g_{ij}$, or $O(\epsilon^2)$
terms in $g_{0i}$.  This has
certainly
been true for all those theories 
analysed by the present author or by others to date,
including several, such as the Rosen bimetric theory,
with $c_g \ne 1$ \citep{lcnw76}.  Such correction terms generically arise from
non-linear terms in the field equations, which are minimally of order
$m/r \sim \epsilon$ higher than the leading term, or from $O(v^2) \sim
O(\epsilon)$
corrections in the definition of the stress-energy tensor.  However,
the retardation effects in the potentials could in principle lead to
such ``half-order'' terms in a slow-motion expansion.  It is precisely these
terms whose effects we wish to explore.

Before proceeding, however, we emphasize that we have assumed a specific
theoretical framework for the analysis.  Relatively few theories with $c_g
\ne c$ have been analysed within the PPN framework (the Rosen theory and a
few others).  Furthermore, it is entirely possible that $c_g \ne c$ theories
may be more fully or consistently realizable in the context of non-metric
theories, for which the PPN framework is not applicable.  In the absence of
a concrete example, we will restrict attention to the PPN approach described
above.

In a metric theory of gravity, the gravitationally coupled Maxwell equations
require that light rays propagate
on null geodesics of the metric (\ref{metric}), given by the equations
\begin{eqnarray}
d^2 x^i/dt^2 + N^\mu N^\nu (\Gamma^i_{\mu\nu} -
N^i \Gamma^0_{\mu\nu})&=&0  \,, 
\nonumber \\
(ds/dt)^2 = g_{00} + 2g_{0i} N^i + g_{ij} N^iN^j &=& 0 \,,
\label{geodesic}
\end{eqnarray}
where $N^\mu = dx^\mu/dt$ ($N^0=1$) along the light ray.

Calculation of the time delay follows the method of \citet{tegp}:
write the trajectory of the light ray in the form
\begin{equation}
x^i(t) = x^i_e + k^i (t-t_e) +  x_p^i (t) \,,
\label{trajectory}
\end{equation}
where $e$ denotes the event of emission of the signal, $k^i$ is a
spatial, Cartesian unit vector, and $x_p^i (t)$ denotes the
perturbation in the path of the ray, which is clearly of order
$\epsilon$.  Then, substituting $N^i = k^i + {\dot x}^i_p$ and
$g_{\mu\nu} = \eta_{\mu\nu} + h_{\mu\nu} $, where $\eta_{\mu\nu}$ is
the Minkowski metric,
into the second of Eqs. (\ref{geodesic}), we obtain
${\bf k} \cdot {\bf \dot x}_p = -h_{\mu\nu} k^\mu k^\nu /2
+O(\epsilon^2)$, where
$k^\mu = (1,k^i)$ and overdot denotes $d/dt$.  
But from Eq. (\ref{trajectory}), we see that 
$|{\bf x} - {\bf x}_e| = t -t_e + {\bf k} \cdot {\bf x}_p +
O[|x_p|^2/(t-t_e)]$,
consequently, 
\begin{eqnarray}
t -t_e &=& |{\bf x} - {\bf x}_e| + \Delta (t,t_e) \,,
\nonumber \\
\Delta (t,t_e) &=& {1 \over 2} k^\mu k^\nu \int_{t_e}^t h_{\mu\nu}
[t,{\bf x}(t)] dt + O(\epsilon^2) |{\bf x} - {\bf x}_e|\,.
\label{Deltat}
\end{eqnarray}
Equation (\ref{Deltat}) is completely equivalent to Eqs. (20) and
(44) of
\citet{KopSch99} and to Eq. (18) of \citet{Kop02}.  
To the required order, the function ${\bf x}(t)$
that is used to evaluate $h_{\mu\nu}$ in the integral is the
unperturbed path of the ray, given by ${\bf x}(t) = {\bf x}_e +
{\bf k} (t -t_e)$.

We now specialize the metric to the physical situation of a 
set of bodies of mass
$m_a$, and velocity ${\bf v}_a$ (the solar system).  
We work in a coordinate system whose
origin is at the barycenter of the system, so that $\sum_a m_a
{\bf v}_a = 0$.  
We parametrize the trajectory of each body by ${\bf x}_a(u)$, where $u$ is a
parameter that is proportional to coordinate time (with unit proportionality
constant), such that 
${\bf x}_a(u=t_0)$ is the location of the body at coordinate time $t=t_0$,
${\bf x}_a(u=0)$ is the location of the body at coordinate time $t=0$,
and so on.  The ordinary velocity is given by ${\bf v}_a(u) = d{\bf
x}_a(u)/du = d{\bf x}_a(u)/dt$.  
(We introduce this parametrization merely to avoid confusion with other
``times'' to be discussed.)
Treating each body as effectively a ``point'' mass, with
$\rho(t^\prime,{\bf x}^\prime) = \sum_a m_a \delta^3 [{\bf x}^\prime
- {\bf x}_a (u=t^\prime)]$ we find that the potentials $U$ and $V_i$ take
the Li\'enard-Wiechert form, 
\begin{eqnarray}
U(t,{\bf x}) &=& \sum_a {m_a \over {|{\bf x} - {\bf x}_a(s_a)| -
{\bf v}_a(s_a) \cdot ({\bf x} - {\bf x}_a(s_a))/c_g}} \,,
\nonumber \\
V_i (t,{\bf x}) &=& \sum_a {m_av_a^i(s_a) \over {|{\bf x} - {\bf x}_a(s_a)| -
{\bf v}_a(s_a) \cdot ({\bf x} - {\bf x}_a(s_a))/c_g}} \,,
\label{UVLW}
\end{eqnarray}
where $s_a$ is ``retarded'' time given by the implicit equation
\begin{equation}
s_a = t - |{\bf x} - {\bf x}_a(s_a)|/c_g \,. 
\label{retarded}
\end{equation}
It is important to understand precisely the meaning of the expressions
${\bf x}_a(s_a)$ and ${\bf v}_a(s_a)$.  They do not mean that ${\bf
x}_a$ and ${\bf v}_a$ have become {\em functions} of $s_a$; they are functions
{\em only} of the parameter $u$.  Rather the expression ${\bf x}_a(s_a)$
means  ${\bf x}_a(u=s_a)$, or
``${\bf x}_a (u)$ evaluated at $u=s_a$ where $s_a$ is
given by evaluating Eq. (\ref{retarded})
for a chosen field point $(t,{\bf x})$''.  This will be important, for
example, when we carry out Taylor expansions of ${\bf x}_a(s_a)$ and
${\bf v}_a(s_a)$.  These issues are addressed in detail in an
Appendix.

Substituting Eqs. (\ref{UVLW}) into Eqs. (\ref{metric}) and thence
into (\ref{Deltat}), we can write the time delay in the form
\begin{equation}
\Delta(t_r,t_e) = (1+\gamma) \sum_a m_a \int_{t_e}^{t_r}
   {{[1-(2+\zeta) {\bf k} \cdot {\bf v}_a(s_a)] dt}
   \over 
    {|{\bf x} - {\bf x}_a(s_a)| -
{\bf v}_a(s_a) \cdot ({\bf x} - {\bf x}_a(s_a))/c_g}} \,,
\label{Delta2}
\end{equation}
where $t_r$ denotes the time of reception of the ray and where $\zeta
= \alpha_1/(2+2\gamma)$.

\section{Calculation of the time delay to 1.5PN order}
\label{lienard}

Equation (\ref{Delta2}) is a sum of contributions for each body in the
system.  Since we are working to linear order in $m$, we can evaluate
the integral for each body separately, then multiply each by $m_a$ and
sum over the bodies.
We rewrite  the unperturbed trajectory of the light ray in the more convenient
form
\begin{equation}
{\bf x}(t) = {\bf k} \sigma + {\bf \xi}_a \,,
\end{equation}
where $\sigma$ is the time $t$ modulo a constant, 
and ${\bf \xi}_a$ is a vector from the barycenter to the light ray at the
moment of its closest approach to body $a$; the parameter $\sigma$ is chosen so
that $\sigma=0$ at this point.
Because we are working to first order in $m_a$, we can ignore all effects
related to the deflection of this ray.

We adopt the simplified notation
\begin{eqnarray}
{\bf r}_a(\sigma,s_a) &=& {\bf x}(\sigma) - {\bf x}_a(s_a) \,,
\nonumber \\
r_a(\sigma,s_a) &=& |{\bf x}(\sigma) - {\bf x}_a(s_a)| \,,
\nonumber \\
s_a &=& \sigma - r_a(\sigma,s_a)/c_g \,,
\label{retardedtime}
\end{eqnarray}
where we have dropped the irrelevant arbitrary constant that
initializes time.
The integral to be evaluated is 
\begin{equation}
\Psi_a(\sigma_r,\sigma_e) = \int_{\sigma_e}^{\sigma_r}
{{[1-(2+\zeta) {\bf k} \cdot {\bf v}_a(s_a)]
d \sigma} \over  {r_a(\sigma,s_a) - {1 \over c_g} {\bf v}_a(s_a) \cdot {\bf
r}_a(\sigma,s_a) }}\,.
\label{psiintegral}
\end{equation}

We now divide the integral into two contributions $\Psi_a^{(1)}$ 
and $\Psi_a^{(2)}$, where
$\Psi_a^{(1)} = \int_{\sigma_e}^{\sigma_1} \dots
d\sigma$ 
and $\Psi_a^{(2)} = \int_{\sigma_1}^{\sigma_r} \dots d\sigma$ 
where the time parameter $\sigma_1$ is chosen
to be large and negative (recall we chose 
$\sigma = 0$ at closest approach), and to
satisfy the following inequalities
$|{\bf x}_a| \ll |\sigma_1| \ll (1/v_a)|{\bf x}_a|$.  
Since $v_a \ll 1$, an appropriate value of $\sigma_1$ clearly exists.  Then
in evaluating $\Psi_a^{(1)}$, we can expand the integrand 
in powers of $1/\sigma$, since $|\sigma|
\gg |{\bf x}_a|$ everywhere, and integrate by parts.  
In evaluating $\Psi_a^{(2)}$,
since the integration is always within the near zone, we can expand the
variables of the bodies about their values at the coordinate time
corresponding to closest approach in a slow-motion
expansion.  We will work only to first order in $v_a$.  Each integral will
depend on the arbitrarily chosen value of $\sigma_1$.  We then match the
integrals by expanding the relevant terms in $\Psi_a^{(1)}$ in a slow-motion
expansion, and the relevant terms in $\Psi_a^{(2)}$ in powers of $1/\sigma_1$.  
We will see that all dependence on $\sigma_1$ cancels to the order
considered, as it must.

We first evaluate $\Psi_a^{(2)}$.  
We re-initialize the parameter $u$ for body $a$ so that it is zero 
at the time of closest approach
and expand the
variables of body $a$ about that point,
so that 
${\bf x}_a(u) = {\bf x}_a(0) + {\bf v}_a(0) u +
O(\epsilon u)$.  
When evaluated at that value of $u$ equal to 
retarded time $s_a$, we get an analogous
result ${\bf x}_a(u=s_a) = {\bf x}_a(0) + {\bf v}_a(0) s_a+
O(\epsilon s_a) $.  
Also, ${\bf v}_a(u=s_a) = {\bf v}_a(0) +
O(\epsilon^{1/2} v_a)$.  Note that ${\bf x}_a(0)$ and ${\bf v}_a(0)$ are
evaluated at $u=0$.  
By defining the vector 
${\bf d}_a(s_a) \equiv {\bf \xi}_a - {\bf x}_a(s_a)$, which
points from the body at retarded time 
to the point of closest approach, we can write
${\bf r}_a(\sigma,s_a) = {\bf k} \sigma + {\bf d}_a(s_a)$.
We note that, to the necessary order, the denominator in 
Eq. (\ref{psiintegral})
can be written $r_a(\sigma,s_a) - {\bf v}_a(s_a) \cdot {\bf
r}_a(\sigma,s_a)/c_g = |{\bf D}| + O(\epsilon r_a)$, where
\begin{eqnarray}
{\bf D} &=& {\bf r}_a(\sigma,s_a) - r_a(\sigma,s_a){\bf v}_a(s_a)/c_g
\nonumber \\
&=& {\bf k} \sigma + {\bf d}_a(s_a)-{\bf v}_a(s_a) (\sigma -s_a) 
\nonumber \\
&=& {\bf k} \sigma + {\bf d}_a(0)-{\bf v}_a(0)\sigma
  + [ {\bf d}_a(s_a)-{\bf d}_a(0) +{\bf v}_a(0)s_a ]
  \nonumber \\
  && + O(\epsilon r_a) \,.
\label{dvector}
\end{eqnarray}
The term in square brackets in Eq. (\ref{dvector}) vanishes to the necessary
order.  Notice that $c_g$ has cancelled out.

Defining the vector ${\bf z}(\sigma) = {\bf k}\sigma + {\bf d}_a(0)$, with
${\bf k} \cdot {\bf d}_a(0) = 0$ ($\bf k$ and ${\bf d}_a$ are orthogonal
at the moment of closest approach to body $a$), we can write 
${\bf D} = {\bf z}(\sigma) - {\bf v}_a(0)\sigma$,  
and  thus
\begin{eqnarray}
r_a(\sigma,s_a) -{\bf v}_a(s_a) \cdot {\bf r}_a(\sigma,s_a)/c_g
&=& | {\bf z}(\sigma) - {\bf v}_a(0)\sigma |
+ O(\epsilon r_a)
\nonumber \\
&=& \bigl [ \sigma^2 (1-2{\bf k} \cdot {\bf v}_a(0))
   + 2\sigma ({\bf k} - {\bf v}_a(0))\cdot {\bf d}_a(0)
   \nonumber \\
   && 
   + d_a(0)^2 \bigr ]^{1/2} + O(\epsilon r_a) \,.
\end{eqnarray}
Inverting this expression and integrating with respect to $\sigma$ yields
the result
\begin{eqnarray}
\Psi_a^{(2)} &=&  \left [ 1 - (1+\zeta){\bf k} \cdot {\bf v}_a(0) \right ]
\ln \left \{ \biggl ( z(\sigma)+{\bf k} \cdot {\bf z}(\sigma) \biggr )
	\left ( 1 - {\bf \hat z}(\sigma) \cdot {\bf v}_a(0)
	 \right ) \right \}_{\sigma_1}^{\sigma_r}  
+ O(\epsilon)\,.
	\label{psi2}
\end{eqnarray}
At the reception point, $\sigma_r$, we define ${\bf z}(\sigma_r) = 
{\bf k}\sigma_r + {\bf \xi} -
{\bf x}_a(0) = {\bf x}_r(\sigma_r)-{\bf x}_a(0) 
\equiv {\bf \tilde x}_{ra}$ and notice
the fact that $d_a(0)^2 = {\tilde r}_{ra}^2 - ({\bf k} \cdot {\bf \tilde
x}_{ra})^2$.  At
$\sigma_1$, we expand the integral in powers of $d_a/|\sigma_1| \ll 1$, using
the expansions (to the required order)
\begin{eqnarray}
z(\sigma_1) &=& -\sigma_1 - d_a(0)^2/2\sigma_1 + 
 d_a(0)^4/8\sigma_1^3 + O(d_a(0)^6\sigma_1^{-5}) \,, 
 \nonumber \\
z(\sigma_1) + {\bf k} \cdot {\bf z}(\sigma_1) &=&
-(d_a(0)^2/2\sigma_1)(1 - d_a(0)^2/4\sigma_1^2 )+  O(d_a(0)^6\sigma_1^{-5}) \,,
 \nonumber \\
{\bf \hat z}(\sigma_1)  &=& - {\bf k} - {\bf d}_a(0)/\sigma_1
  + {\bf k} (d_a(0)^2/2\sigma_1^2) + O(d_a(0)^3\sigma_1^{-3}) \,.
\label{zexpand}
  \end{eqnarray}
Substituting these expansions into Eq. (\ref{psi2}), and keeping terms through 
$O(\sigma_1^{-2})$, we obtain
\begin{eqnarray}
\Psi_a^{(2)} &=& -  \biggl [ 1 - (1+\zeta) {\bf k} \cdot {\bf v}_a(0)
 \biggr ]
\ln \left ({{{\tilde r}_{ra} - {\bf k} \cdot {\bf \tilde x}_{ra}} \over 2}
\right ) -  {{{\bf \tilde x}_{ra} \cdot {\bf v}_a(0)} \over 
	{{\tilde r}_{ra} }} 
- {\bf k} \cdot {\bf v}_a(0) 
\nonumber \\
&&
+ \biggl [ 1 - (1+\zeta) {\bf k} \cdot {\bf v}_a(0)
 \biggr ]
\ln |\sigma_1| + {{d_a(0)^2} \over {4 \sigma_1^2}}
\nonumber \\
&&
- {{{\bf d}_a(0) \cdot {\bf v}_a(0)} \over {\sigma_1}}
+ \left ({1-\zeta \over 4} \right ) 
{{d_a(0)^2 {\bf k} \cdot {\bf v}_a(0)} \over 
	{ \sigma_1^2}} + O(\epsilon,d_a(0)^3\sigma_1^{-3}) \,.
	\label{V2final}
\end{eqnarray}

Next, we evaluate $\Psi_a^{(1)}$.  
Here, because $\sigma$ is integrated over a range
that could include many dynamical periods of the masses as the signal
propagates from the distant star toward the near zone, 
we must leave all variables such as ${\bf x}_a(s_a)$
unexpanded.  Instead, we expand all expressions in advance
in powers of $1/\sigma$.  Then
\begin{eqnarray}
{\bf r}_a(\sigma,s_a) &=& {\bf k}\sigma + {\bf d}_a(s_a) \,,
\nonumber \\
r_a(\sigma,s_a) &=& -\sigma - {\bf k} \cdot {\bf d}_a(s_a)
	- {{d_\perp (s_a)^2} \over {2\sigma}}
	+ {{{\bf k} \cdot {\bf d}_a(s_a) d_\perp (s_a)^2} \over {2\sigma^2}}
	+ O(d_a(s_a)^4\sigma^{-3}) \,,
\nonumber \\
s_a &=& \left ( 1+ {1 \over c_g} \right ) \sigma + {{{\bf k} \cdot {\bf
d}_a(s_a)} \over c_g} +  {{d_\perp (s_a)^2} \over {2c_g\sigma}}
	- {{{\bf k} \cdot {\bf d}_a(s_a) d_\perp (s_a)^2} 
\over {2c_g\sigma^2}}
\nonumber \\
&&+ O(d_a(s_a)^4\sigma^{-3}) \,,
\label{sigmaexpand}
\end{eqnarray}
where $d_\perp(s_a) = d_a(s_a)^2 - ({\bf k}\cdot {\bf d}_a(s_a))^2$.  
Calculating $[r_a(\sigma,s_a) - {\bf v}(s_a) \cdot 
{\bf r}_a(\sigma,s_a)/c_g]^{-1}$, and expanding it in powers of 
$\sigma^{-1}$, and then expanding to first order in $\epsilon^{1/2}$, 
we obtain the integral 
\begin{eqnarray} 
\Psi_a^{(1)} &=& - \int_{\sigma_e}^{\sigma_1} d\sigma  \biggl \{ 
{1 \over \sigma} 
\left ( 1 - \eta {{{\bf k} \cdot {\bf v}_a(s_a)} \over c_g} \right )
	\nonumber \\
	&&
	-  {1 \over \sigma^2} \left ( {\bf k} \cdot {\bf d}_a(s_a) 
	 + {{{\bf d}_a(s_a) \cdot {\bf v}_a(s_a)} \over c_g}
	 - 2\eta^\prime 
	{{{\bf k} \cdot {\bf d}_a(s_a) {\bf k} \cdot {\bf v}_a(s_a)} 
	 \over c_g} \right ) 
\nonumber \\
&& - {1 \over \sigma^3} \left ( {1 \over 2} [d_a(s_a)^2 - 3({\bf k}\cdot
{\bf d}_a(s_a))^2] - {{{\bf k}\cdot {\bf v}_a(s_a)} \over c_g}
[\eta^\prime d_a(s_a)^2 - 4\eta^{\prime\prime}({\bf k}\cdot {\bf d}_a(s_a))^2]
\right .
\nonumber \\
&&
\left .
- 2{{{\bf k}\cdot {\bf d}_a(s_a) {\bf k} \cdot{\bf v}_a(s_a)} \over c_g}
\right )
+ {1 \over 2\sigma^4} \left ( 3d_a(s_a)^2 {\bf k}\cdot {\bf d}_a(s_a)
	-5 ({\bf k}\cdot {\bf d}_a(s_a))^3 \right ) \biggr \}
	\nonumber \\
	&& 
	+ O(\epsilon,d_a(s_a)^4 \sigma^{-4}) \,,
\end{eqnarray}
where $\eta=1+(2+\zeta)c_g/c$, $\eta^\prime=(1+\eta)/2$ and
$\eta^{\prime\prime}= (5+3\eta)/8$.
We then integrate the various terms by parts, using the formula $\int (d\sigma /
\sigma^{n+1}) f(s_a) = -f(s_a)/n\sigma^n + \int (d\sigma /n\sigma^n)
df/d\sigma$.  Each $\sigma$-derivative $df/d\sigma$
raises the order of that term by $\epsilon^{1/2}$; we stop integrating
by parts when the resulting term is $O(\epsilon)$ relative to the
leading term.  Because $\sigma$ in the integral
parametrizes the light path, and because we are not Taylor expanding the
orbital variables, we must take into account that $s_a$ depends on $\sigma$
through Eq. (\ref{sigmaexpand}), so that 
\begin{eqnarray}
{d \over d\sigma} {\bf d}_a(s_a) &=& -{\bf v}_a(s_a) {ds_a \over d\sigma}
\nonumber \\
&=&  -{\bf v}_a(s_a) \left [ \left ( 1 + {1 \over c_g} \right ) 
	- {d_\perp(s_a)^2 \over {2c_g\sigma^2}} \right ] 
	+ O(\epsilon,d_a(s_a)^3\sigma^{-3}) \,.
\end{eqnarray}
After integration, we 
discard terms that depend on negative powers of the emission time
$\sigma_e$; for light from a distant quasar, these are clearly negligible.
The result is
\begin{eqnarray}
\Psi_a^{(1)} &=& - 
\biggl [ 1 -(1+\zeta){\bf k} \cdot {\bf v}_a(s_1) \biggr ]
\ln |\sigma_1| 
+ {1 \over c} 
\biggl [ 1 - (1+\zeta) {\bf k} \cdot {\bf v}_a(s_e) \biggr ]
\ln |\sigma_e|
\nonumber \\
&&
- {1 \over \sigma_1} {\bf k} \cdot {\bf d}_a(s_1)
- {1 \over 4\sigma_1^2} [{d}_a(s_1)^2 - 3 ({\bf k} \cdot {\bf d}_a(s_1))^2 ]
\nonumber \\
&&
+{1 \over 6\sigma_1^3} [3{d}_a(s_1)^2 {\bf k} \cdot {\bf d}_a(s_1)
	-5 ({\bf k} \cdot {\bf d}_a(s_1))^3 ]
\nonumber \\
&&
+ {1  \over 2\sigma_1} \left [ 
	 \left ( 1 - {1 \over c_g} \right ) {\bf d}_a(s_1)
	\cdot {\bf v}_a(s_1)
	+ \left ( 1+2\zeta + {1 \over c_g} \right ) {\bf k} \cdot {\bf d}_a(s_1)
	{\bf k} \cdot {\bf v}_a(s_1) \right ]
\nonumber \\
&&
- {1  \over 4\sigma_1^2} \left [ 
	 2 \left ( 1 - {1 \over c_g} \right )  {\bf k} \cdot {\bf d}_a(s_1)
	 {\bf d}_a(s_1) \cdot {\bf v}_a(s_1)
	 - (1+\zeta) {d}_a(s_1)^2 {\bf k} \cdot {\bf v}_a(s_1)
	 \right .
	 \nonumber \\
	 &&
	 \left .
	 +  \left ( 1+3\zeta + 2 {1 \over c_g} \right )
	 {\bf k} \cdot {\bf v}_a(s_1) ({\bf k} \cdot {\bf d}_a(s_1))^2
	 \right ]
	+ O(\epsilon,d_a(s_a)^4\sigma_1^{-4})
	\,.
	\label{V1}
\end{eqnarray}
Since $\sigma_1$ is within the near zone, we can use the expansions 
\begin{eqnarray}
{\bf d}_a(s_1) &=& {\bf d}_a(0) - {\bf v}_a(0) s_1 +O(\epsilon d_a)
\nonumber \\
&=& {\bf d}_a(0) - {\bf v}_a(0) \left [
 \left ( 1 + {1 \over c_g} \right ) \sigma_1
	 + {d_a(0)^2 \over {2c_g\sigma_1}} \right ]
		 + O(\epsilon,d_a(0)^4\sigma_1^{-3})
\nonumber \\
{\bf v}_a(s_1) &=& {\bf v}_a(0) + O(\epsilon) \,,
\end{eqnarray}
where, again, $(0)$ means $u=0$.
Recalling that ${\bf k} \cdot {\bf
d}_a(0)=0$, and substituting these expressions
into Eq. (\ref{V1}), we obtain
\begin{eqnarray}
\Psi_a^{(1)} &=& 
 \biggl [ 1 - (1+\zeta) {\bf k} \cdot {\bf v}_a(s_e)  \biggr ]
 \ln |\sigma_e| + {\bf k} \cdot {\bf v}_a(0) 
 \left ( 1 + {1 \over c_g} \right )
 \nonumber \\
 &&
 -
 \biggl [ 1 - (1+\zeta) {\bf k} \cdot {\bf v}_a(0)  \biggr ]
 \ln |\sigma_1|
- {{d_a(0)^2} \over {4 \sigma_1^2}}
\nonumber \\
&&
+ {{{\bf d}_a(0) \cdot {\bf v}_a(0)} \over {\sigma_1}}
- \left ({1-\zeta \over 4} \right ) {{d_a(0)^2 {\bf k} \cdot {\bf v}_a(0)} \over
	{\sigma_1^2}} + O(\epsilon,d_a(0)^3\sigma_1^{-3}) \,.
\label{V1final}
\end{eqnarray}
Combining the expressions (\ref{V2final}) and (\ref{V1final})
for $\Psi_a^{(2)}$ 
and $\Psi_a^{(1)} $, we see that all terms involving $\sigma_1$ cancel, as
expected.  Our final expression for $\Psi_a$ is
\begin{eqnarray}
\Psi_a &=& -
   \biggl [ 1 - (1+\zeta)  {\bf k} \cdot {\bf v}_a(0) \biggr ]
 \ln \left ({{{\tilde r}_{ra} - {\bf k} \cdot {\bf \tilde x}_{ra}} \over 2}
 \right ) - {{{\bf \tilde x}_{ra} \cdot {\bf v}_a(0)} \over
	 {{\tilde r}_{ra} }}
	 -{{{\bf k} \cdot {\bf v}_a(0)} \over c_g }
	 \nonumber \\
	 &&
+  \biggl [ 1 -  (1+\zeta)  {\bf k} \cdot {\bf v}_a(s_e) \biggr ]
  \ln |\sigma_e| + O(\epsilon) \,.
\label{Vfinal}
\end{eqnarray}

Notice that the dependence on $c_g$ in the term 
${\bf k} \cdot {\bf v}_a(0) /c_g$ in
Eq. (\ref{Vfinal}) is illusory: to
obtain the measured effect, we must multiply $\Psi_a$ for body $a$ by its mass
$m_a$, and sum over all the bodies in the system.  Although the event
corresponding to $\sigma=0$ (closest approach) 
is different for each body, the effect of these
differences on $v_a$ 
is of order of the acceleration of each body, which is $O(\epsilon)$, 
hence we can evaluate ${\bf v}_a(0)$
at the same coordinate time for each body, say, the time of reception
of the ray.  But, because $\sum_a m_a {\bf
v}_a = 0$ in barycentric coordinates, the net effect of this
$c_g$-dependent term  vanishes.  
Similarly, for a given light ray, the time $\sigma_e$ is a constant,
independent of the body.  The difference between the various values of $s_e$
for different bodies has an effect only at order $\epsilon$, so again,
summing over all  the bodies in the system causes the velocity correction to
the $\ln |\sigma_e|$ term to vanish.  The $\ln |c\sigma_e|$ term then
is an irrelevant constant.  By the same argument, the factor of 2 in
the logarithm produces only an irrelevant constant.
The final formula for the time delay,
correct to 1.5PN order is
\begin{equation}
\Delta(t_r,t_e) = - (1+\gamma) \sum_a m_a \biggl \{
	\biggl [ 1-(1+\zeta) {\bf k} \cdot {\bf v}_a(0) \biggr ]
 \ln \left ({\tilde r}_{ra} - {\bf k} \cdot {\bf \tilde x}_{ra}
 \right ) + {{{\bf \tilde x}_{ra} \cdot {\bf v}_a(0)} \over
         {{\tilde r}_{ra} }} \biggr \} 
\,.
\label{semifinal}
\end{equation}
The speed of propagation of gravity, which initially
appeared in the retarded potentials, is
nowhere to be found in this final expression.  The only consequence of using
an alternative theory of gravity is in the PPN parameters $\gamma$ and
$\zeta=\alpha_1/(2+2\gamma)$.  

The variable ${\bf \tilde x}_{ra} = {\bf x}(\sigma_r)-{\bf x}_a(0)$
can be converted so that all times are referred to the moment of
reception of the ray, using the expressions
\begin{eqnarray}
{\bf \tilde x}_{ra} &=& {\bf x}_{ra} +
{\bf v}_a \sigma_r 
\nonumber \\
&=& {\bf x}_{ra} +
{\bf v}_a {\bf k} \cdot {\bf \tilde x}_{ra} \,,
\nonumber \\
{\bf v}_a(0) &=& {\bf v}_a + O(\epsilon) \,,
\end{eqnarray}
where ${\bf x}_{ra}
\equiv {\bf x}(\sigma_r)-{\bf x}_a(\sigma_r)$ 
and ${\bf v}_a \equiv {\bf v}_a(\sigma_r)$.  Substituting into
Eq. (\ref{semifinal}) and rearranging, we find, to the necessary
order,
\begin{equation}
\Delta(t_r,t_e) = - (1+\gamma) \sum_a m_a \biggl \{
        \biggl [ 1-(1+\zeta) {\bf k} \cdot {\bf v}_a \biggr ]
 \ln \left (r_{ra} - {\bf K} \cdot {\bf x}_{ra} \right )\biggr \}
\,,
\label{final}
\end{equation}
where
\begin{equation}
{\bf K} = {\bf k} - {\bf k} \times ({\bf v}_a \times {\bf k}) \,.
\end{equation}
In general relativity ($\gamma=1$, $\zeta=0$) this agrees completely
with results derived by \citet{KopSch99}, to 1.5PN
order.

\section{Discussion}
\label{discussion}

We have shown that the speed of propagation of gravity has no
direct influence on the time delay of light to 1.5PN order.
The only effect comes from any modification of the PPN parameters
that might arise in a theory with a different propagation speed.  This
contradicts claims made by \citet{Kop01,Kop02}.   

In addition, existing solar-system experiments already place sufficiently
strong bounds on the PPN parameters $\gamma$ and $\alpha_1$ that even the
potential difference from GR in the 1.5PN correction
represented by the parameter $\zeta$ in Eq.
(\ref{final}) must be small.  The parameter $\gamma$
is known to be unity to a part in 1000 from standard time delay and light
deflection measurements, while $|\alpha_1| < 2 \times 10^{-4}$ 
from analyses of
Lunar laser ranging \citep{muller} and binary pulsar data \citep{bell}.
The latter bounds assume that the solar system and the relevant binary
pulsar are moving with respect to the cosmic background radiation with known
velocities of order 300 km/s, and that, in any theory of gravity with
$\alpha_1 \ne 0$, the rest frame of that radiation coincides with a
cosmological preferred frame.  Such a frame is to be expected in a theory of
gravity in which $c_g \ne c$ for the following reason: in a theory with
either interacting dynamical fields or with non-dynamical
``prior-geometrical'' fields, the speed of gravity is presumably a function
of some cosmological values of the fields in the theory.  But because $c_g
\ne c$, its value depends on the velocity of the frame in which it is
measured.  There must therefore exist some frame in which $c_g$ takes its
value directly 
from the cosmologically induced values of the underlying fields -- the
only logical frame is the mean rest frame of the universe, and hence the
rest frame of the CBR.  Then, if $\alpha_1 \ne 0$, there will be measurable
effects in any system that moves relative to this preferred frame.
With these strong bounds, $|\zeta| < 5 \times 10^{-5}$.

Nevertheless a weaker bound on $\alpha_1$ can be
inferred from experiments without appealing to preferred-frame effects.
We assume only that the theory is a
Lagrangian-based theory  of gravity, so that it
possesses suitable conservation laws for total momentum and energy.
Then,
Lunar laser ranging tests of the Nordvedt effect bound the combination
$|4\beta-\gamma-3-10\xi/3-\alpha_1-2\alpha_2/3| < 10^{-3}$. 
The perihelion advance of Mercury combined with the existing bounds on
$\gamma$ yield $|\beta -1| < 3 \times 10^{-3}$; 
Earth tide measurements force the
bound $|\xi| < 10^{-3}$.   One is then left with the relatively generous
bound $|\alpha_1-2\alpha_2/3| < 1.6 \times 10^{-2}$.  
All bounds on $\alpha_2$ come from considerations of preferred-frame
effects.  Thus without appealing to preferred-frame tests, 
one would have to have an
extraordinary
cancellation between $\alpha_1$ and $\alpha_2$ to conform to
Nordtvedt effect measurements, while still making  
$\zeta$ large enough to make a measurable difference in Eq.
(\ref{final}).   

\acknowledgments

We are grateful to Steve Carlip, Sergei Kopeikin, Gerhard Sch\"afer, Kenneth Nordtvedt,
and Hideki Asada for useful discussions.
This work is supported in part by the National Science Foundation,
under grant no. PHY00-96522.

\appendix

\section{Meaning of retarded time}

Throughout these calculations, quantities such as ${\bf x}_a(s_a)$ and ${\bf
v}_a(s_a)$ appear, where $s_a$ is retarded time evaluated using Eq.
(\ref{retarded}).  In the near zone, we use the expansion ${\bf
x}_a(s_a) = {\bf x}_a(0) + {\bf v}_a(0) s_a + \dots$.  One is tempted
to state that the $(0)$ that appears means something other than 
coordinate time equals zero.  But this would not be
correct.  The trajectory of the body ${\bf x}_a(u)$ is a function
only of the parameter $u$, which is chosen to be directly proportional to
coordinate time.  By contrast, $s_a$ is not an
independent variable, instead it is a two-point dependent variable,
determined by both the field point $(t,{\bf x})$ and the body's
trajectory.  Equation (\ref{retarded}) serves to determine (albeit
implicitly or iteratively) that value of the parameter $u=s_a$ at which to
evaluate ${\bf x}_a(u)$.  The equation quoted above is then simply
a special case of the Taylor expanded equation ${\bf
x}_a(u) = {\bf x}_a(0) + {\bf v}_a(0) u + \dots$, evaluated at
$u=s_a$.  Figure 1 illustrates this explicitly.  On the
$t^\prime,x^\prime$ plane, assuming flat spacetime, the world line of
a body moving with fixed velocity $v$ is shown, together with a field
point at $(t,x)$.  The point on the trajectory that corresponds to
retarded time from the field point is given by the intersection
$\cal P$
between the particle trajectory and the past null ray from the field point (we
assume for this illustration that the speed of the null ray is
unity).  This occurs at a time $|x-x_a(s_a)|$ earlier than the field
time $t$, therefore it occurs at a time $t^\prime =
t-|x-x_a(s_a)| \equiv s_a$, which thus corresponds to the value $u=s_a$
along the trajectory.  By considering the projection of this event
onto the $x^\prime$-axis, together with the location and 
velocity of the particle
at $t^\prime=0$, it is clear that $x(u=s_a) = x(0)+v(0)s_a$.

\clearpage

\begin{figure}
\plotone{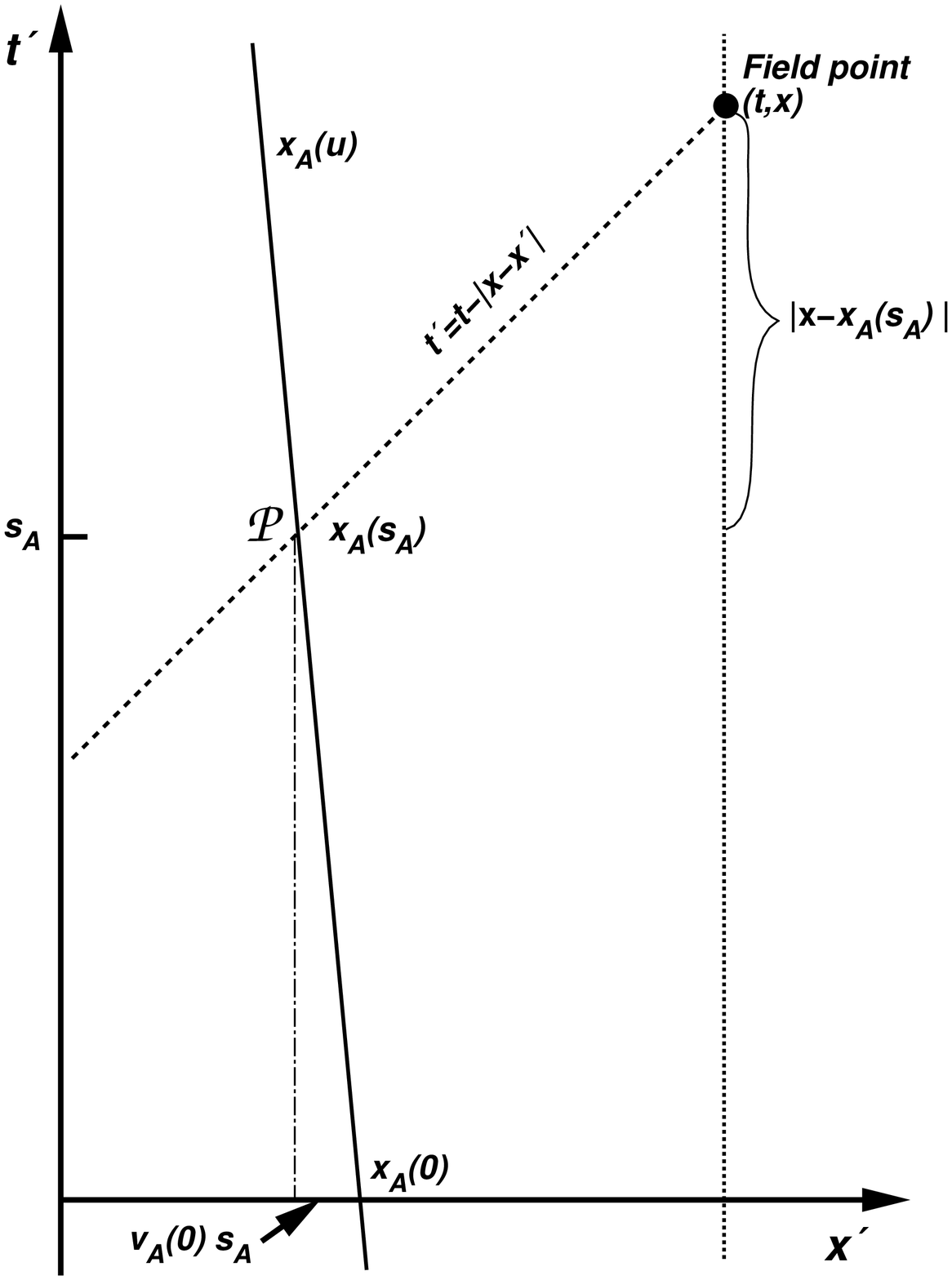}
\caption{Illustration of retarded time}
\end{figure}

\end{document}